## *Simulation of the phononless hopping in a Coulomb glass*

J. Matulewski[*,1,2], S. D. Baranovskii[2], **and** P. Thomas[2]

[1] Instytut Fizyki, Uniwersytet Mikołaja Kopernika, ul. Grudziądzka 5, 87-100 Toruń, Poland
[2] Fachbereich Physik und Zentrum für Materialwissenschaften der Philipps-Universität, Marburg Renthof 5, 35032 Marburg/Lahn, Germany

**PACS** 72.20.Ee, 71.30.+h

The phononless hopping conductivity of a disordered system with localized states is studied in a broad range of frequencies by straightforward computer simulations taking into account Coulomb interactions. At sufficiently low temperatures, the conductivity is determined by the zero-phonon absorption of the photon by pairs of states. The laser frequency dependence of the conductivity is examined and compared with the analytical model of Efros and Shklovskii and with recent experimental data obtained on Si:P. The range of parameters is determined, for which the conductivity dependence on photon energy best reproduces the experimental results.

**1. Introduction**
Recently Helgren et al. [1] and Lee et al. [2,3] studied the frequency-dependent conductivity of Si:P at low temperature. The results were interpreted as indication for phononless hopping in the electron glass regime. (The properties of the electron glass were reviewed, for instance, in Ref. [4].) The experimental data show a transition from linear to a quadratic dependence of the conductivity with increasing photon energy (laser frequency), which is interpreted by the authors as a transition from Coulomb glass behavior determined by the long range Coulomb interactions, to Fermi glass behavior, in which these interactions are less important. This crossover was predicted already in the early 80's by Efros and Shklovskii (ES) [5,6]. For photon energies larger than the mean Coulomb interaction energy of the pairs, the ES theory reproduces the quadric Mott-like frequency dependence of the conductivity [7].
The crossover visible in Helgren's experiment is even stronger than predicted by ES theory. Also other experiments performed recently (for references see [2]) show that the ES theory though being qualitatively correct needs some quantitative improvements. The aim of this paper is to check if it is possible to reproduce the results of the above-mentioned experiments in computer simulations of the Coulomb glass system, which does not suffer from approximations necessary in the analytical description.

**2. Model**
The model which we simulate, is represented by a set of $n$ identical donors and a set of $Kn$ identical acceptors. All donors have equal energies of ground states if isolated (here set to zero) and equal localization lengths $\alpha$. Nevertheless, with the Coulomb interactions taken into account, the donor energies are not equal since donors and acceptors are randomly distributed in space and they interact with each other via Coulomb forces that induce shifts of donor energies. One of the characteristic features of this system in the ground state is the so-called Coulomb gap in the single-particle density of states (SP-DOS), which is located around the Fermi energy [6].
Elfros and Shklovskii calculated the conductivity of such a system induced by laser light at zero temperature in the pair approximation and derived the famous analytic formula [5]:
$$\sigma(\omega) = \omega\, r^4(\omega)(\hbar\omega + 1/r(\omega))\,,\quad (1)$$
where $r(\omega)$ is the most probable space distance (frequency-dependent) of sites with the energy difference (including the Coulomb term) equal to the photon energy of the absorbed light $\Gamma = \hbar\omega$, that is
$$r(\omega) = \alpha \ln(2I_0/\hbar\omega)\,.\quad (2)$$

---

[*] Corresponding author: e-mail: Jacek.Matulewski@Physik.Uni-Marburg.DE

In Eq. (2) $I_0$ is the pre-exponential factor of the overlap of the wave functions of two states participating in the electron transition [5]. According to the formula (1) in a limit of low laser frequencies, i.e. for photon energy $\hbar\omega$ lower than the energy of the mean Coulomb interaction $1/r(\omega)$, the conductivity depends linearly on the laser frequency $\omega$. For $\hbar\omega > 1/r(\omega)$ the frequency dependence becomes quadratic, in agreement with the result of Mott for a non-interacting Fermi glass [7]. In other words according to ES theory, the presence of the Coulomb interaction influences the conductivity only in the low frequency limit and the crossover from the Coulomb glass to the Fermi glass behavior occurs approximately for photon energies equal to the mean energy of the Coulomb interaction between the sites of the absorbing pair.

The assumption of a distinct relation between the spatial distance and the energy difference $\Gamma$ of donors in the pair which contributes to the conductivity given by Mott's law, i.e. Eq. (2), is a weak point of analytical theories. In order to avoid this approximation we carry out numerical simulations. It allows us to verify both the analytical theory and the interpretation of the experiments given by Helgren and his colleagues.

We study below the properties of the so-called pseudoground states that are the states with the lowest possible energy, which can be reached from the initial, usually randomly chosen, configuration by one electron transitions [8,9].

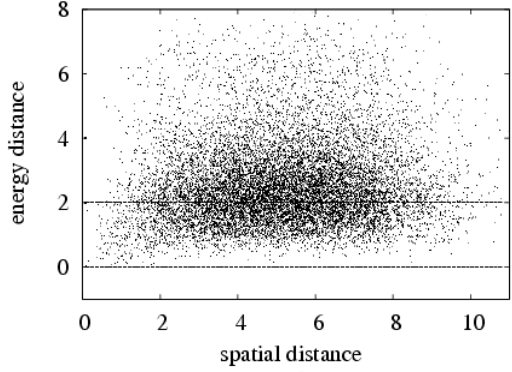

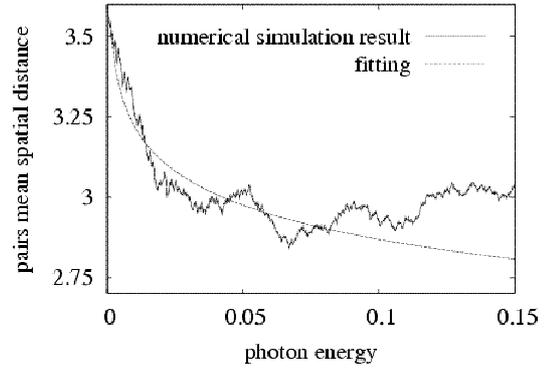

**Fig. 1**   The set of pairs contributing to the conductivity process presented in the coordinates of spatial and energy distance for $T=0$, $N=500$, $\alpha=0.27$.

**Fig. 2**   Dependence of the pairs mean spatial distance on photon energy for $T=0$, $\alpha=0.27$, $N=1000$, $K=0.5$; the number of Monte-Carlo steps is 1000.

### 3. Results and discussion

In order to adjust the simulations to the experimental conditions [1] we have chosen the following values of the material parameters. The concentration is 69% of the critical concentration which for Si:P is $n_C = 3.52 \cdot 10^{24}$ m$^{-3}$. This means that $n = 2.4288 \cdot 10^{24}$ m$^{-3}$. In dimensionless units (defined by taking $e^2 n^{1/3}/\kappa$ as an energy and temperature unit and $n^{-1/3}$ as a length unit) the value of the localization length depends on the concentration. One can assume a reasonable value of the localization length $\alpha$ for Si:P lying between 20 Å and 30 Å [10]. This means that its value in dimensionless units lies between 0.27 and 0.36. Notice that for concentration equal to 50% of $n_C$ the value 20 Å of the localization length corresponds in dimensionless units to 0.24. Thus fitting the concentration parameter is not so important since the accurate value of the localization length is not known. The compensation ratio $K=0.5$ was assumed in the simulations.

Calculating of the pseudoground state of the whole system for zero temperature, which is the most time consuming step of our simulations, gives us explicit information about all donor energies and allows a direct calculation of the distribution of pairs and the conductivity averaged over realizations. Fig. 1 shows the obtained results in one realization set of pairs contributing to the conductivity, presented in coordinates of spatial and energy distances. The most probable energy difference for sites creating the pairs that contribute to the conductivity is 2. This is because for $T=0$ it is most probable to find an occupied donor at the left and an empty one at the right maximum of the SP-DOS (not shown). For the parameters used to prepare Fig. 1, pairs with energy differences equal to about 2 are usually spatially separated by 4 to 7 units. It corresponds to the wide area with the larger density of points representing the pairs in Fig. 1. Examining the pairs with energy smaller than 3, one can notice that their mean distance also decreases until the energy difference (or the frequency of the absorbed photon) reaches the value of about 0.1. If we further decrease the laser frequency, the pairs mean spatial distance $r(\omega)$ restricted to the particular frequency will rise (see Fig. 2). The dependence of the mean spatial distance between the sites in a pair on the energy difference can be fitted by the Mott's formula given by Eq. (2). The fitting parameter is the site wave-function-overlap pre-exponential factor $I_0$. For the localization length equal to $\alpha=0.27$ the fitted value of $I_0$ is about $3 \cdot 10^4$ and for $\alpha=0.36$ it is $7 \cdot 10^3$. In both cases the fitting was performed for dimensionless energies from 0.001 to 0.1. Despite the ability of fitting the mean value of $r(\omega)$ obtained from the simulation by Mott's law, one should be aware of the important difference between these two quantities (especially in the context of making use of it in ES's derivation of formula (1)). Mott's law says that there is a distinct dependence of the value of the spatial distance of sites in the pair involved in conductivity for the given frequency of the laser while the simulation results presented in Fig. 1 convince us that such distinct dependence is not justified and the distribution of the spatial distance $r(\omega)$ is wide also for low frequencies $\omega$. It even broadens for $\hbar\omega<0.05$ what is the direct reason of the growth of mean value of spatial distance in this range. It is of course a great advantage of the numerical simulations that no such assumptions are necessary and more reliable and precise results may be obtained.

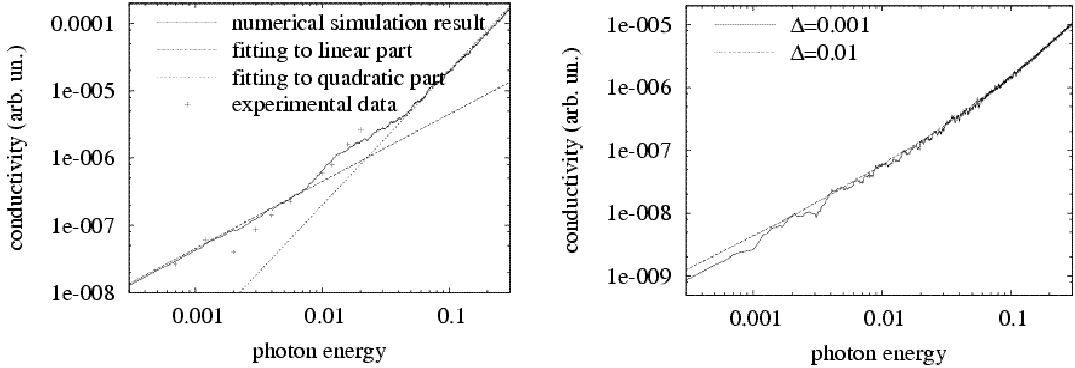

**Fig. 3** Scaled to $\Delta$ conductivity as a function of frequency at zero temperature for $\alpha=0.36$ (left) and $\alpha=0.27$ (right). Other parameters are $N=1000$, $K=0.5$. Each point is an average over 25000 realizations.

With the set of pairs obtained from the first stage of the simulation one can calculate the conductivity of the randomly disordered system of donors and acceptors incorporating the Coulomb interaction of sites in the presence of the low frequency laser field. The results are shown in Fig. 3 for two values of the localization length. For $\alpha=0.36$ (left plot) one can see the crossover from linear to quadric dependence on photon energy, which is stronger then that predicted by ES, but quite well reproduces the results of the experiment performed by Helgren et al. [1]. The crossover point ($\omega \approx 0.01 \approx 500$ GHz) agrees with that seen in the experimental data (it is approximately the value of the mean Coulomb interaction between sites in the system). The crossover for $\alpha=0.27$ (right plot) is much more smooth and broad. The conductivity for $\alpha=0.27$ is plotted for two values of the width of the laser frequency $\Delta$ (that is the range of frequencies in which the conductivity is summed up while creating the histogram) and scaled to it. One can see that it converges already for quite a large value of $\Delta$. The results in Fig. 3 (left) suggest a complex structure of the conductivity dependence on the laser frequency. The second crossover is visible for $\hbar\omega$ equal to about 0.05.

A comparison with experiment is the crucial point, since the accurate value of the localization length α and the compensation $K$ of the samples used in the experiment are not known precisely and our numerical calculations indicates its importance. Since we try to relate our results to the experiment, we have performed calculations for the parameters taken from [1], namely for 69% Si:P as it was mentioned above. We assumed the localization length value for this type of material to be in the interval between 0.27 and 0.36. After checking the wider range of α one can determine that the crossover from the linear to the quadric dependence similar to the one presented in Fig. 3 (left) is most clearly visible for the localization values between 0.35 and 0.40 (for $K$=0.5 and $T$=0). Outside this range it is weaker like in Fig. 3 (right). For α<0.15 one cannot determine the crossover point at all, while the distributions of the pairs which take part in the conductivity have always a clearly visible crossover at the photon energy equal to about $\hbar\omega$=0.01. Nevertheless for α>0.15 the value of the conductivity is not strongly dependent on the localization range (not shown).

## 4. Conclusions

The AC conductivity in the Coulomb glass in the absence of phonons was examined using the numerical method based on the pair approximation. The numerical approach allows us to avoid approximations that are necessary in order to obtain the analytical results. In particular, we compare our result with the recent measurements of the conductivity in the Si:P samples. It was possible to determine the localization length α=0.36 (in units $n^{-1/3}$) for which it is possible to reproduce the experimental value of the photon energy for which the transition from linear to quadric dependence of conductivity occurs. Also the shape of the conductivity dependence on the photon energy is reproduced quite well in our simulations. The crossover visible at the experiment and in our results (for α=0.36) is stronger than that predicted by analytical theories.

**Acknowledgements**  Financial support of the Deutsche Forschungsgemeinschaft, of the European Graduate College "Electron-Electron Interactions in Solids" Marburg-Budapest and that of the Fonds der Chemischen Industrie is gratefully acknowledged.